# The design of eutectic high entropy alloys in Al-Co-Cr-Fe-Ni system

Ali Shafiei

Metallurgy Group, Niroo Research Institute (NRI), Tehran 14665-517, Iran
E-mail: alshafiei@nri.ac.ir, Tel: +98 (21) 88074187

**Abstract**

In the present work, a simple approach is proposed for predicting the compositions of eutectic high entropy alloys (EHEAs) in Al-Co-Cr-Fe-Ni system. It is proposed that eutectic lines exist between certain eutectic alloys in this system and, as a result, new eutectic or near-eutectic compositions can be obtained by mixing the alloys which are located on the same eutectic line. The approach is applied for a series of experimentally verified eutectic alloys and new eutectic or near-eutectic alloys are designed for Al-Co-Cr-Fe-Ni system. Furthermore, by investigating the compositions of verified eutectic alloys in Al-Co-Cr-Fe-Ni system, compositional diagrams are proposed which show the relations between the concentrations of constituent elements in eutectic alloys. The compositional diagrams suggest that EHEAs are derived from binary and ternary eutectic alloys. Moreover, the proposed diagrams can be considered as convenient methods for evaluating the composition of EHEAs in Al-Co-Cr-Fe-Ni system.

## 1. Introduction

High entropy alloys (HEAs) are a new group of metallic alloys which have attracted significant attentions because they could have desirable properties [1-5]. These alloys contain multiple (at least five) principal elements in difference with traditional alloys which are usually based on one principal element [1-5].

A subgroup of HEAs are eutectic high entropy alloys (EHEAs) which are reported recently by Lu et al. [6]. Due to their fine in-situ lamellar composite microstructures, EHEAs have shown very promising combinations of strength and ductility [6-15] which have encouraged materials scientists to focus on EHEAs as a promising new class of structural alloys. Furthermore, these alloys may be considered as fillers for brazing due to their low melting temperatures and good viscosity. How easy to design EHEAs is still a challenge. That is because the phase diagrams are not available for quaternary and quinary alloy systems. In fact, most of the EHEAs are found by a trial and error approach. Some simple techniques are proposed for predicting the compositions of EHEAs [16-18], but the developed approaches can mostly predict near eutectic compositions [16-18]. Therefore, the objective of the present work is proposing a simple method for designing new EHEAs. Al-Co-Cr-Fe-Ni system is considered in this work; but it is believed that the approach which is proposed here may also be used for other alloy systems.

## 2. Methodology

### Alloy preparation and characterization

The ingot of alloy $Al_{19.5}Co_{47.5}Cr_{7.5}Ni_{25.5}$ was prepared via arc melting the high purity constituent elements (Ni(99.9%), Co(99.99%), Al(99.999%), and Cr (99.9%)) under a Ti-gettered high purity argon atmosphere. The ingot was remelted four times to achieve compositional homogeneity. By using a water-cooled copper mold, the homogenized ingot was suction casted into a 4 cm long and 8 mm diameter rod. The sample was further sectioned perpendicular to its length for microstructural investigations which were performed by an optical microscope. Marble's etchant was used for the etching of the sample.

### Thermodynamic simulations

All of the simulations and thermodynamic calculations in the present work are performed by JMatPro® software version 7.0.0 developed by Sente Software Ltd. [19]. The software uses a multicomponent thermodynamic database, NiData, developed by ThermoTech Ltd. [19]. “Nickel-based superalloy” toolbox is used for the simulations. The ability of the software for modeling the solidification behaviors of various multi-component alloys is verified by several research works [20-24]. An alloy is considered as eutectic if simulation results predict that the solidification occurs in a narrow temperature range ($\Delta T_{max}$=10 °C), and if during the solidification simultaneous formation of two solid phases γ (fcc) and B2 (bcc) occurs. This criterion is considered here according to the obtained simulation results for verified EHEAs [25].

## 3. Model development

The chemical compositions of some of the verified binary, ternary, quaternary and quinary eutectic alloys in Al-Co-Cr-Fe-Ni system are shown in Table 1. Only binary and ternary eutectic alloys with eutectic phases γ (FCC) and B2 (BCC) are considered here. So the model which is proposed here is only valid for γ + B2 eutectic alloys. All of the reported quaternary and quinary EHEAs in Al-Co-Cr-Fe-Ni system also form phases γ and B2 during their eutectic reactions (Table 1). For ternary systems the binary eutectic reaction (L→ γ + B2) occurs along a line (Figure 1) [26-30], therefore ranges of compositions are listed for these alloys in Table 1. The eutectic lines for ternary systems are shown in Figure 1. The eutectic lines are in fact not entirely straight according to the ternary phase diagrams; however, for simplicity, it is assumed here that eutectic lines are straight as they are shown in Figure 1. For Al-Co-Ni, Al-Fe-Ni and Al-Cr-Ni systems the liquidus projections are available [26-30], so eutectic lines are directly extracted from phase diagrams (Figures 1a-1c) [26-30]. However, for ternary Al-Co-Fe and Al-Co-Cr systems liquidus projections are not available. The eutectic lines for these systems are obtained by following procedures. For Al-Co-Fe system the minimum eutectic composition is reported to be $Al_{15}Co_{63}Fe_{22}$ at% [31-32]. Therefore, it is assumed that a straight eutectic line exists between the eutectic composition $Al_{20}Co_{80}$ [34] and $Al_{15}Co_{63}Fe_{22}$ [31-32] (Figure 1d). For Al-Co-Cr system neither a liquidus projection nor a minimum eutectic composition is reported. However, Liu et al. [33] modeled the isothermal section of Al-Co-Cr system at 1300 °C, and from the isothermal section, the boundaries of the two phase region γ+B2 can be obtained (Figure 1e). According to the boundaries of the two phase region γ+B2 at 1300 °C, the eutectic line could be predicted which is shown in Figure 1e. It should be noted that for both Al-Co-Fe and Al-Co-Cr systems the eutectic lines may not be entirely straight, but straight eutectic lines according to Figure 1 are accurate enough for developing the model.

Table 1. The chemical compositions of some of the experimentally verified binary, ternary, quaternary and quinary eutectic alloys in Al-Co-Cr-Fe-Ni system

| **Alloy** | **Chemical composition (at.%)** | | | | | Eutectic phases |
|---|---|---|---|---|---|---|
| | Al | Co | Cr | Fe | Ni | |
| AlCo [34] | 20 | 80 | - | - | - | γ + B2 |
| AlCoFe [31-32] | 15-20 | 63-80 | - | 0-22 | - | γ + B2 |
| AlFeNi [26] | 17-22 | - | - | 10-50 | 33-68 | γ + B2 |
| AlCoNi [29-30] | 20-23 | 7-80 | - | - | 0-70 | γ + B2 |
| AlCoCr [33] | 19-20 | 56-80 | 0-25 | - | - | γ + B2 |
| AlCrNi [27-28] | 17.5-23 | 0 | 4-31 | 0 | 51.5-73 | γ + B2 |
| AlCoFeNi [13] | 19 | 20 | - | 20 | 41 | γ + B2 |
| AlCoCrNi [12] | 19 | 15 | 15 | - | 51 | γ + B2 |
| AlCoCrNi [14] | 17.4 | 21.7 | 21.7 | 0 | 39.2 | γ + B2 |
| AlCoCrNi [14] | 16 | 38.6 | 22.7 | 0 | 22.7 | γ + B2 |
| $AlCrFeNi_2$ [35] | 16 | - | 20 | 20 | 44 | γ + B2 |
| AlCoCrFeNi [36] | 18 | 30 | 10 | 10 | 32 | γ + B2 |
| $AlCoCrFeNi_{2.1}$ [6] | 16.4 | 16.4 | 16.4 | 16.4 | 34.4 | γ + B2 |
| $AlCo_2CrFeNi_2$ [18] | 17 | 28.6 | 14.3 | 14.3 | 25.8 | γ + B2 |
| $AlCoCrFe_2Ni_2$ [18] | 17 | 14.3 | 14.3 | 28.6 | 25.8 | γ + B2 |
| $AlCoCrFeNi_3$ [18] | 17 | 14.3 | 14.3 | 14.3 | 40.1 | γ + B2 |
| AlCoCrFeNi [37] | 18 | 30 | 10 | 10 | 32 | γ + B2 |
| AlCoFeNi [37] | 18 | 30 | - | 20 | 32 | γ + B2 |
| AlCoFeNi [37] | 18 | 27.34 | - | 27.34 | 27.34 | γ + B2 |
| AlCoCrFeNi [37]* | 18 | 30 | 10 | 10 | 30 | γ + B2 |
| AlCoCrFeNi [37]* | 18 | 24 | 10 | 10 | 36 | γ + B2 |
| AlCoCrFeNi [37]* | 18 | 20 | 10 | 10 | 40 | γ + B2 |
| AlCoCrFeNi [25] | 16 | 41 | 15 | 10 | 18 | γ + B2 |

* These alloys contained 2 at.% of W for improving the mechanical properties

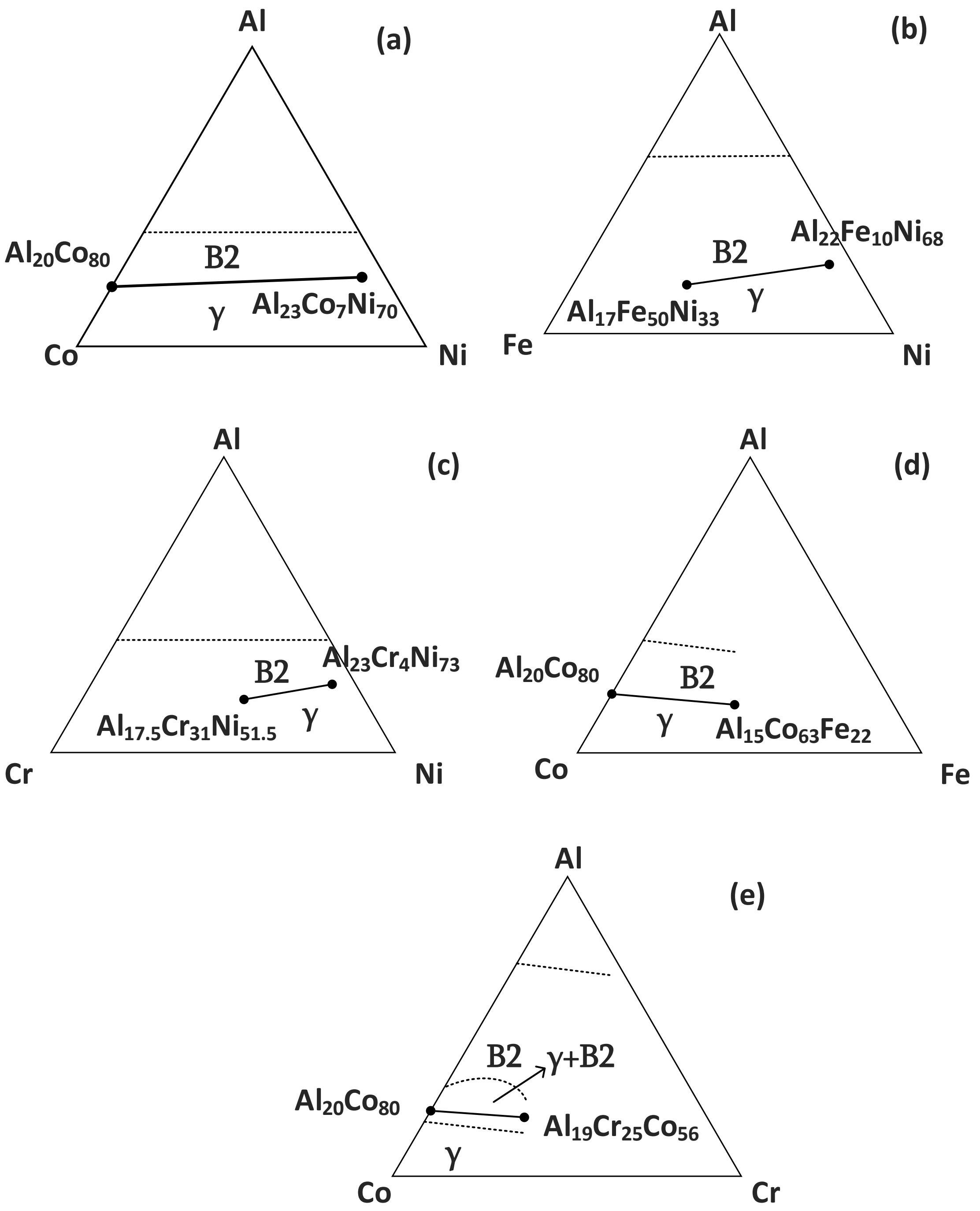

Figure 1. The eutectic lines for (a) Al-Co-Ni [29-30], (b) Al-Fe-Ni [26], (c) Al-Cr-Ni [27-28], (d) Al-Co-Fe, and (e) Al-Co-Cr and systems. For Al-Co-Fe system the eutectic line is drawn between the minimum eutectic composition $Al_{15}Co_{63}Fe_{22}$ [31-32] and eutectic composition $Al_{20}Co_{80}$ [34]. For Al-Co-Cr system the eutectic line is drawn according the boundaries of γ + B2 region at 1300 °C [33]

As it can be seen in Figure 1, Al-Co-Ni, Al-Co-Cr and Al-Co-Fe eutectic lines origin from binary eutectic $Al_{20}Co_{80}$. For example, it can be said that Al-Co-Ni eutectics are formed by adding Ni to binary eutectic $Al_{20}Co_{80}$. Furthermore, it can be supposed that Al-Co-Cr and Al-Co-Fe eutectics are formed by adding respectively Cr and Fe to $Al_{20}Co_{80}$. So, it can be concluded that ternary eutectics Al-Co-Cr, Al-Co-Fe and Al-Co-Ni are originated from binary eutectic $Al_{20}Co_{80}$. Similarly, one can assume that quaternary and quinary EHEAs are originated from binary eutectic $Al_{20}Co_{80}$. Furthermore, ternary eutectic compositions (all of the compositions on the eutectic lines in Figure 1) can be considered as initial compositions for quaternary and quinary eutectic alloys. *In general, it can be assumed that the quaternary and quinary eutectic alloys in Al-Co-Cr-Fe-Ni system are derived from binary and ternary eutectic alloys, and, as a result, it can be concluded that eutectic lines may exist between certain binary, ternary, quaternary and quinary eutectic alloys in Al-Co-Cr-Fe-Ni system*. Based on this assumption, a network or graph structure can be proposed for eutectic alloys in Al-Co-Cr-Fe-Ni system. The proposed network or graph structure is schematically shown in Figure 2. Each line which is shown in Figure 2 is in fact showing a eutectic line which its existence can be verified by experiments or simulations. Therefore, some of the lines in Figure 2 are shown by dashed lines meaning that their existence are not verified by experiments yet. The eutectic network which is shown in Figure 2 could be used as a guideline for designing new eutectic alloys. The application of this network for designing new eutectic alloys is explained in next section. The eutectic network which is shown in Figure 2 only contains a limited number of eutectic alloys and can be expanded by considering all of the eutectic composition in Al-Co-Cr-Fe-Ni.

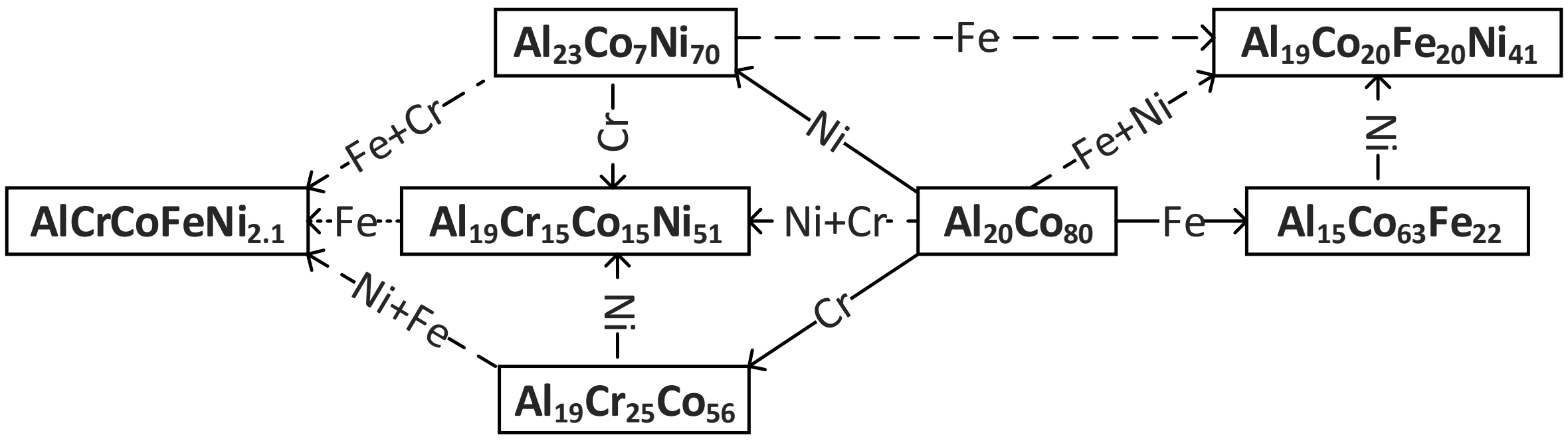

Figure 2. Proposing a network or graph structure for some of the binary, ternary, quaternary and quinary eutectic alloys in Al-Co-Cr-Fe-Ni system; each line in the network represents a eutectic line and dashed lines show eutectic lines which are not verified by experiments; as an example, the above network shows that eutectic $Al_{23}Co_7Ni_{70}$ is made by adding Ni to $Al_{20}Co_{80}$

The concept of eutectic lines can also be explained by considering the phase rule. According to the phase rule ($F = C - P + 1$), a binary eutectic reaction (e. g. L→ γ + B2) has zero degree of freedom ($F$=2-3+1=0) in a binary system. So this eutectic reaction is an invariant reaction in a binary phase diagram. In other words, a binary eutectic reaction can only occur at one temperature and composition in a binary system. In a ternary system, $F$ is equal to 1 ($F$=3-3+1=1) for a binary eutectic reaction. As a result, there is a range of compositions for a binary eutectic reaction in a ternary system; this range is shown by a eutectic line on a ternary phase diagram (Figure 1). Thus, a eutectic line on a ternary phase diagram is a set of eutectic compositions. In other words, each point on a eutectic line represents a eutectic composition. In a quinary system, $F$ is equal to 3 ($F$=5-3+1=3) for a binary eutectic reaction and it can be expected that the eutectic compositions form a volume within the phase diagram. As a result, eutectic lines can also be assumed for eutectic alloys in a quinary system. This is schematically shown in Figure 3. Therefore, a eutectic line for a quinary system is a set of eutectic compositions within the eutectic volume of a quinary phase diagram. This concept is used in the present work, and it is shown that eutectic lines exist between certain eutectic alloys in Al-Co-Cr-Fe-Ni system; the compositions of experimentally verified eutectic alloys are used for finding these eutectic lines.

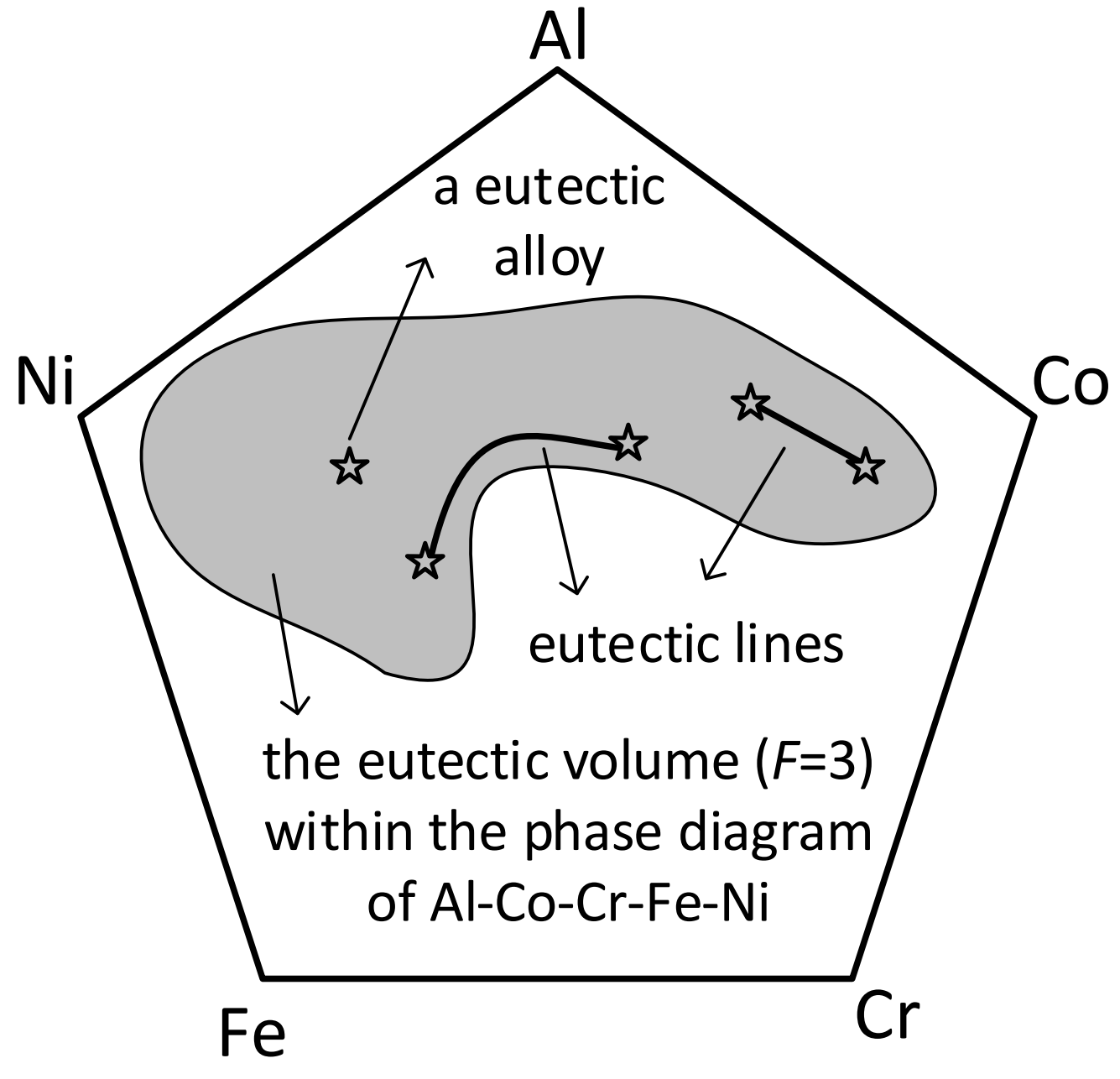

Figure 3. The concept of eutectic lines between eutectic alloys in quinary Al-Co-Cr-Fe-Ni system

## 4. Designing new eutectic alloys

If a eutectic network similar to Figure 2 can be developed for eutectic alloys in Al-Co-Cr-Fe-Ni system, then predicting new eutectic compositions will be very simple and straightforward. As an example, let's assume that a eutectic line exists between eutectic compositions $Al_{20}Co_{80}$ [34] and $Al_{19}Co_{15}Cr_{15}Ni_{51}$ [12] (Figure 4a). Therefore, the alloy $Al_{19.5}Co_{47.5}Cr_{7.5}Ni_{25.5}$ which is at the central point of the eutectic line may be eutectic as well. The simulation result for this alloy is shown in Figure 4b which clearly shows that alloy $Al_{19.5}Co_{47.5}Cr_{7.5}Ni_{25.5}$ is eutectic. Furthermore, the optical images from the microstructure of alloy $Al_{19.5}Co_{47.5}Cr_{7.5}Ni_{25.5}$ are shown in Figure 5. It can be seen that the microstructure of alloy $Al_{19.5}Co_{47.5}Cr_{7.5}Ni_{25.5}$ consists from a fine intimate mixture of two phases indicating that alloy $Al_{19.5}Co_{47.5}Cr_{7.5}Ni_{25.5}$ is eutectic. So the approach was successful in predicting a new eutectic alloy.

As another example, let's assume that a eutectic line exists between eutectic alloys $Al_{23}Co_{7}Ni_{70}$ [30] and $Al_{19}Co_{15}Cr_{15}Ni_{51}$ [12] (Figure 4a). Therefore, the alloy $Al_{21}Co_{11}Cr_{7.5}Ni_{60.5}$ which is at the central point of the eutectic line may be eutectic as well. The simulation results for alloy $Al_{21}Co_{11}Cr_{7.5}Ni_{60.5}$ is shown in Figure 4c which indicates that this alloy is eutectic. Therefore, it may be concluded that a eutectic line exists between two verified eutectic alloys $Al_{23}Co_{7}Ni_{70}$ [30] and $Al_{19}Co_{15}Cr_{15}Ni_{51}$ [12] and new eutectic alloys can be designed by mixing the alloys at various molar ratio.

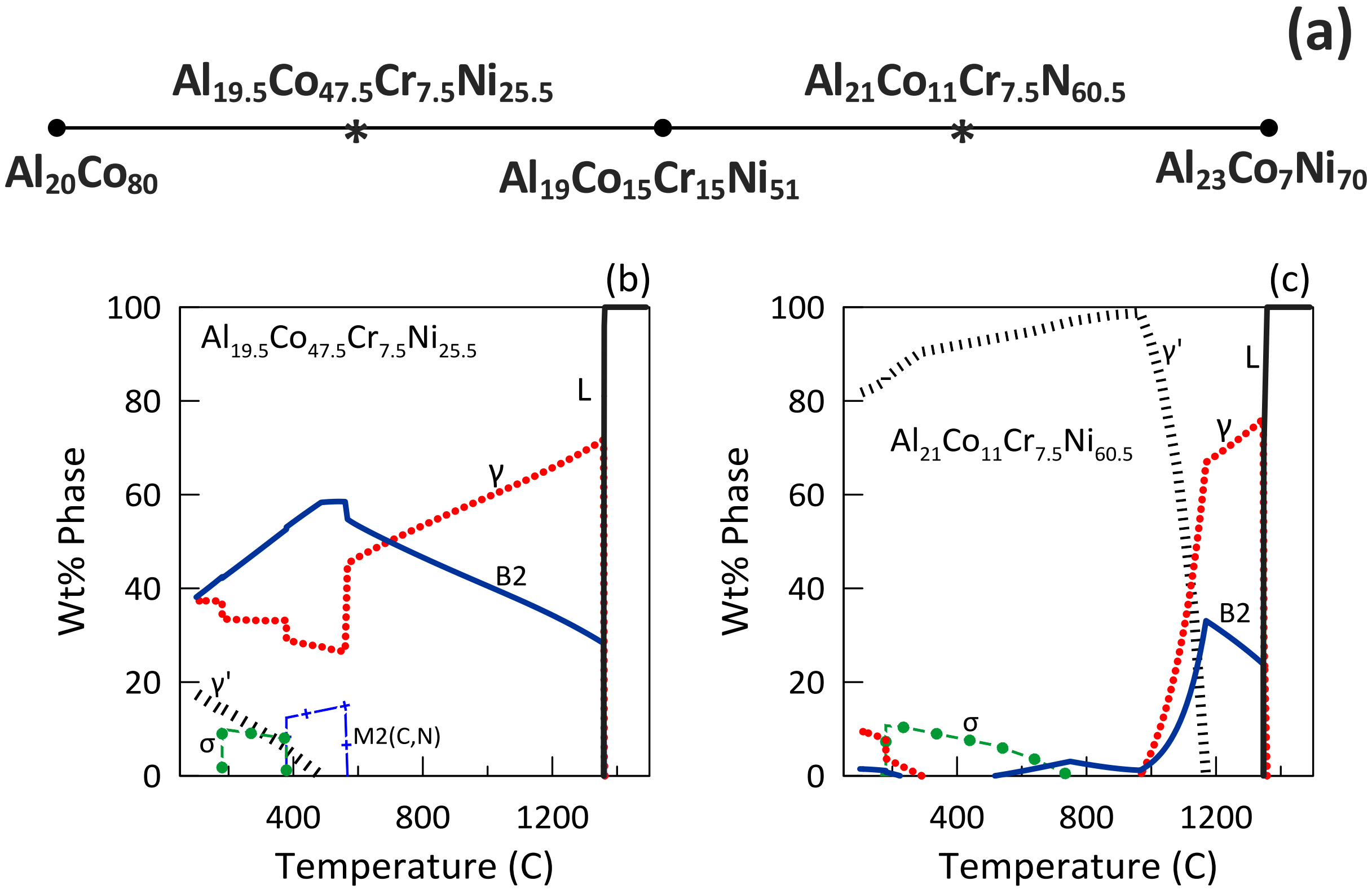

Figure 4. (a) Assuming eutectic lines between eutectic alloys $Al_{20}Co_{80}$ and $Al_{19}Cr_{15}Co_{15}Ni_{51}$ [12] and between $Al_{23}Co_7Ni_{70}$ [30] and $Al_{19}Co_{15}Cr_{15}Ni_{51}$ [12]; obtaining new eutectic alloys with compositions $Al_{19.5}Co_{47.5}Cr_{7.5}Ni_{25.5}$ and $Al_{21}Co_{11}Cr_{7.5}Ni_{60.5}$; the simulation results for alloys (b) $Al_{19.5}Co_{47.5}Cr_{7.5}Ni_{25.5}$ and (c) $Al_{21}Co_{11}Cr_{7.5}Ni_{60.5}$

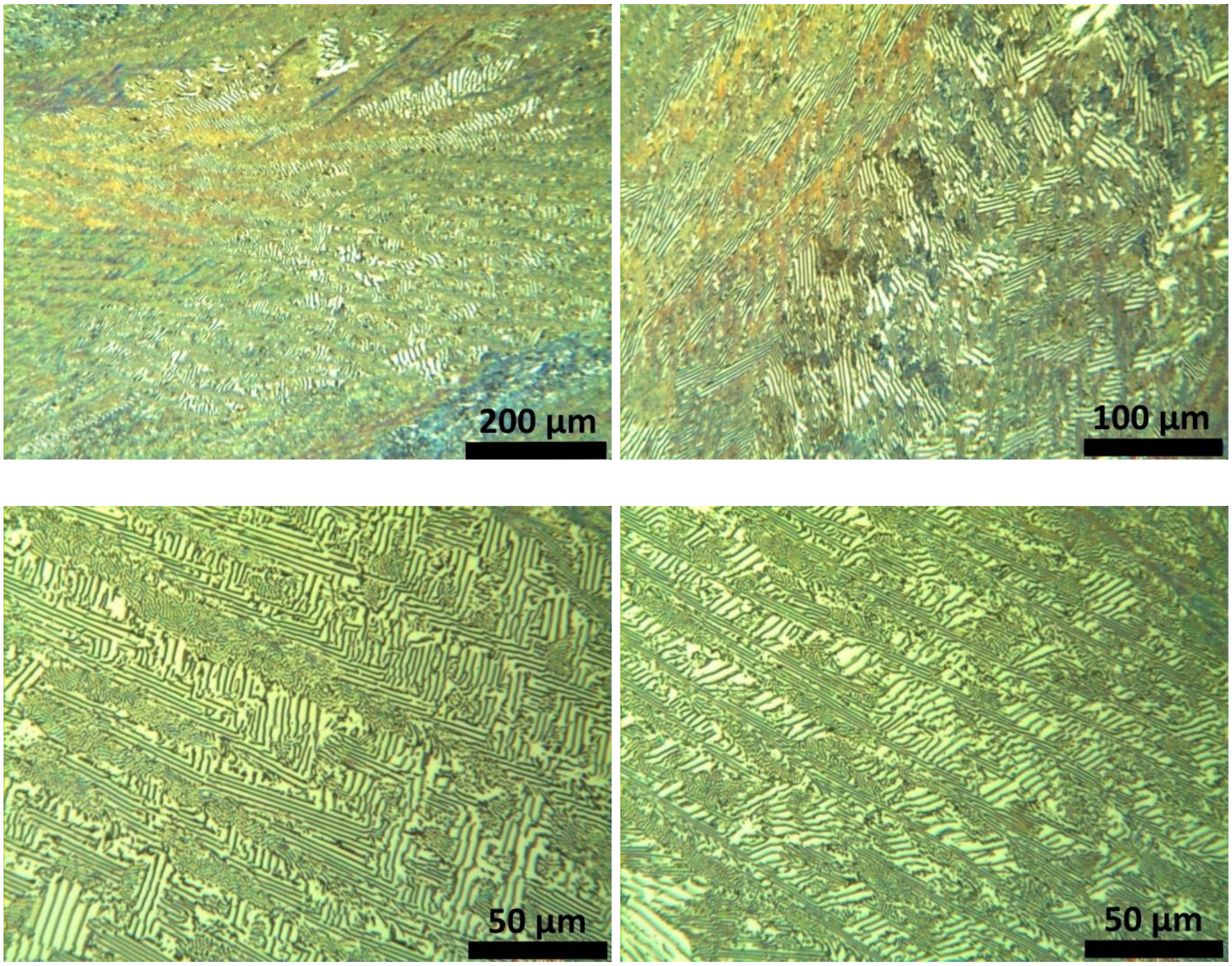

Figure 5. Optical images of the as-cast microstructure of alloy $Al_{19.5}Co_{47.5}Cr_{7.5}Ni_{25.5}$

It should be noted that to accurately confirm that a eutectic line exists between two eutectic compositions, all of the alloy combinations (at 0.01:0.99 to 0.99:0.01 molar ratios) should be examined, and just examining the alloy at the central point may not be enough for confirming the existence of a eutectic line. But, examining all of the alloy combinations is very time consuming. Therefore, in the present work, three alloy combinations (alloys at 0.25:0.75, 1:1 and 0.75:0.25 molar ratios) are examined for investigating the existence of a eutectic line. If it can be shown that these three alloys are eutectic, then the chance for other alloys along that line to be eutectic is probably very high. So, in the present work it is assumed that if mixed alloys at 0.25:0.75, 1:1 and 0.75:0.25 molar ratio are eutectic, then there is a eutectic line between two eutectic alloys.

The results in Figures 4 and 5 show that the assumption of the existence of eutectic lines between eutectic alloys is valid. To further examine this idea, more eutectic alloy combinations from Table 1 are examined by a procedure similar to Figure 4, and three alloy combinations between two eutectic alloys (alloys at 0.25:0.75, 1:1 and 0.75:0.25 molar ratio) are examined. It is assumed that if alloys at 0.25:0.75, 1:1 and 0.75:0.25 molar ratios are eutectic, then there is a eutectic line between two eutectic alloys. The obtained simulation results are shown in Figure 6 as a eutectic network. The narrow continues lines show the eutectic lines which are verified by simulation results. The bold lines show eutectic lines which are verified by experiments (ternary eutectic lines in phase diagrams), and the dotted lines show lines which are examined by the software but are not eutectic according to simulation results. According to the proposed network in Figure 6, it can be seen that some of eutectic alloys are connected via eutectic lines meaning that those alloys may be mixed with each other for making other eutectic alloys. So, Figure 6 provides a guideline for designing new eutectic alloy in Al-Co-Cr-Fe-Ni system. It should be noted that eutectic network which is shown in Figure 6 is made by considering a limited number of eutectic alloy combinations. For making the complete eutectic network, all of the eutectic alloy combinations from Table 1 should be considered.

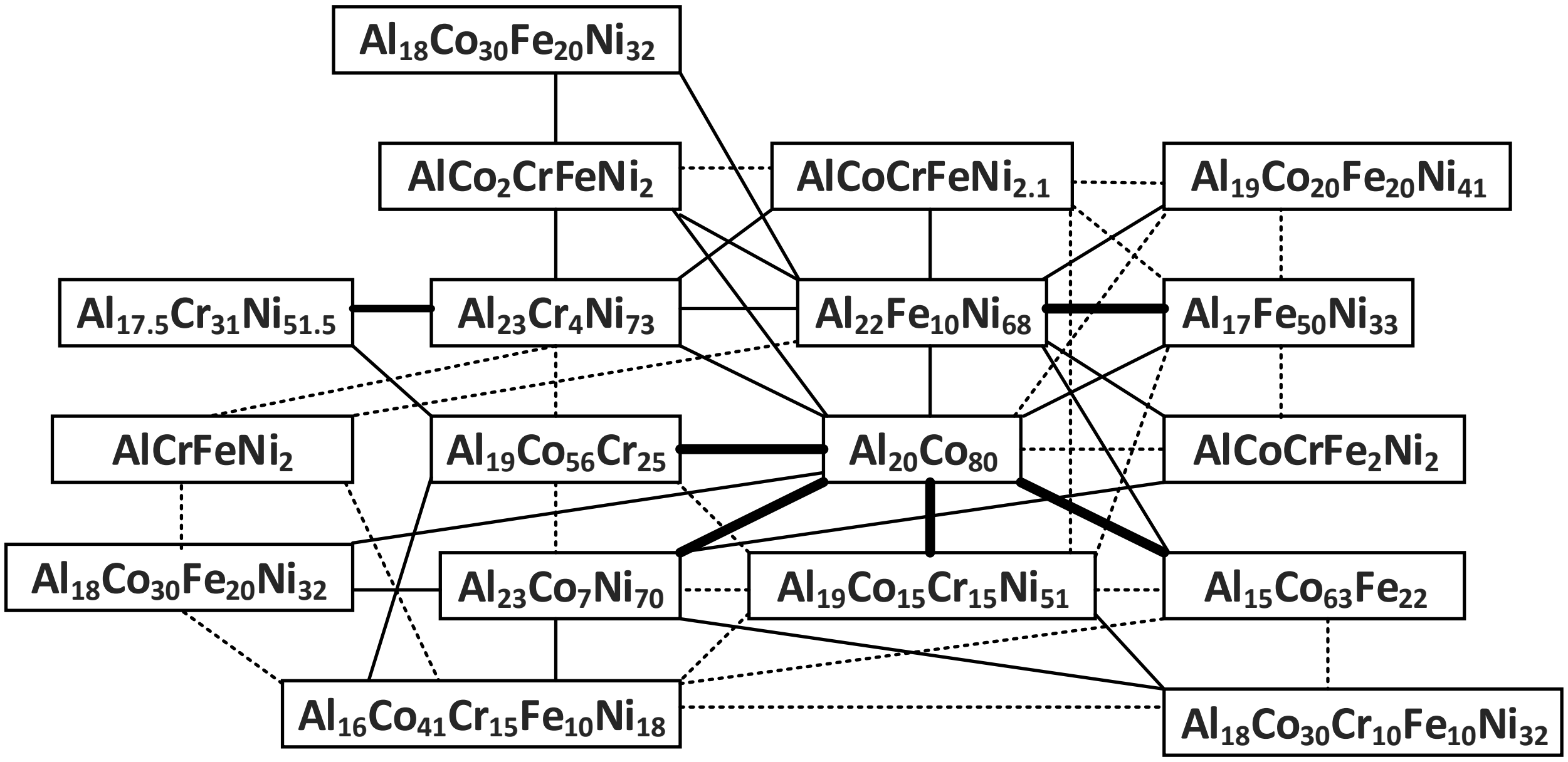

Figure 6. The eutectic network obtained by considering some of the combinations of eutectic alloys in Table 1; the narrow eutectic lines are verified by simulation results, and the bold eutectic lines are verified by experiments, the dotted lines show the lines which are investigated but are not eutectic according to simulation results; the simulation results show that alloys along the dotted lines are in fact near eutectic alloys

The dotted lines in Figure 6 show lines which are examined by software but are not eutectic lines. It is observed that alloys along dotted lines are in fact near eutectic (hypo-eutectic or hyper-eutectic) alloys. Therefore, the alloys which are connected by dotted lines could be used for designing near eutectic (hypo-eutectic or hyper-eutectic) alloys. For example consider eutectic alloys $Al_{22}Fe_{10}Ni_{68}$ [26] and $AlCrFeNi_2$ ($Al_{16}Cr_{20}Fe_{20}Ni_{44}$) [35] which are connected by a dotted line according to Figure 6. By mixing the alloys at 1:1 molar ratio, alloy $Al_{19}Cr_{10}Fe_{15}Ni_{56}$ can be obtained. The simulation results for alloy $Al_{19}Cr_{10}Fe_{15}Ni_{56}$ is shown in Figure 7a which shows that this alloy is a near eutectic alloy because its solidification do not occur in a narrow range of temperature. Therefore, the alloys which are connected by dotted lines can be used for obtaining near eutectic compositions. A curved eutectic line which is schematically shown in Figure 7b can be assumed between alloys $Al_{22}Fe_{10}Ni_{68}$ [26] and $AlCrFeNi_2$ ($Al_{16}Cr_{20}Fe_{20}Ni_{44}$) [35]. According to Figure 7b, when a curved eutectic line (where the eutectic compositions are arranged along a curve not a straight line) exists between two eutectic alloys, then the mixed alloys will not be located on the eutectic line. So the mixed alloys will not be eutectic.

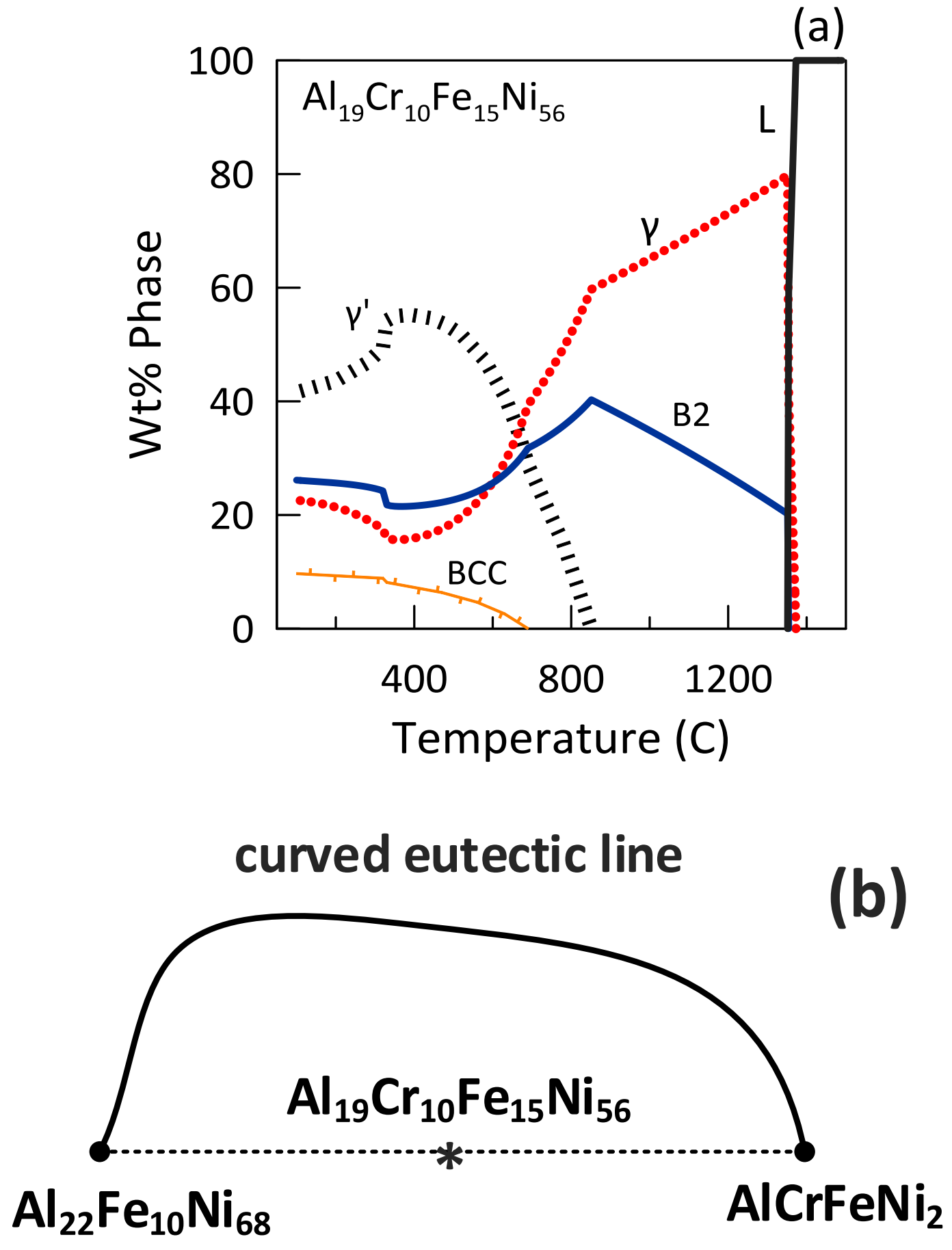

Figure 7. Simulation results for alloy $Al_{19}Cr_{10}Fe_{15}Ni_{56}$ obtained by mixing eutectic alloys $Al_{22}Fe_{10}Ni_{68}$ [26] and $AlCrFeNi_2$ [35], (b) assuming a curved eutectic line between alloys $Al_{22}Fe_{10}Ni_{68}$ [26] and $AlCrFeNi_2$ [35]; in this condition, alloys cannot be mixed with each other for designing new eutectic alloys

The eutectic network which is proposed in Figure 6 is obtained only by thermodynamic simulations, and thermodynamic simulations are indeed used successfully by many research works for predicting the microstructures of alloys [38-43]. So the results in Figure 6 may be valid and applicable. Nevertheless, the most reliable way for examining the existence of eutectic lines between eutectic compositions is performing the experiments. So experiments are needed for accurately confirming the eutectic network in Figure 6. In the present work, just one eutectic composition ($Al_{19.5}Co_{47.5}Cr_{7.5}Ni_{25.5}$) at the central point between alloys $Al_{20}Co_{80}$ [34] and $Al_{19}Co_{15}Cr_{15}Ni_{51}$ [12] is confirmed, but it is believed that more eutectic alloys could be obtained by using a similar approach. In the following section some compositional diagrams are proposed for evaluating the composition of EHEAs in Al-Co-Cr-Fe-Ni system. They could be used alongside with the developed method for evaluating the composition of EHEAs in Al-Co-Cr-Fe-Ni system.

## 5. Compositional diagrams

Because it is proposed that EHEAs are derived from binary and ternary eutectic compositions, therefore it can be expected that the concentration of elements in EHEAs should be within specific ranges. For example, according to Table 1, the Al concentration of binary and ternary eutectic alloys is in the range of 15-23 at.%. Therefore, the Al concentration of EHEAs should be within this range as well. As it can be seen, the Al concentration of EHEAs (Table 1) is indeed within this range. The Al concentration of EHEAs alloys can also be investigated by Figures 8a and 8b. Figure 8a shows the Al concentration versus (Ni+Co) concentration, and Figure 8b shows the Al concentration versus (Cr+Fe) concentration of all eutectic alloys in Table 1. By considering the relation Al+Co+Cr+Fe+Ni=100, these two diagrams have the same meaning; but both diagrams are plotted so the relations could be more easily understood. It can be seen that there are regions limited by binary and ternary eutectics in which all EHEAs are located. This observation suggests that EHEAs are all originated from binary and ternary eutectic alloys. Furthermore, it can be postulated that all EHEAs in Al-Co-Cr-Fe-Ni system should be located within eutectic regions in Figure 8. Moreover, it can be proposed that if the Al concentration of an alloy is not within the eutectic regions in Figure 8, then that alloy cannot be eutectic. So eutectic regions in Figure 8 provide references for evaluating the Al concentration of EHEAs.

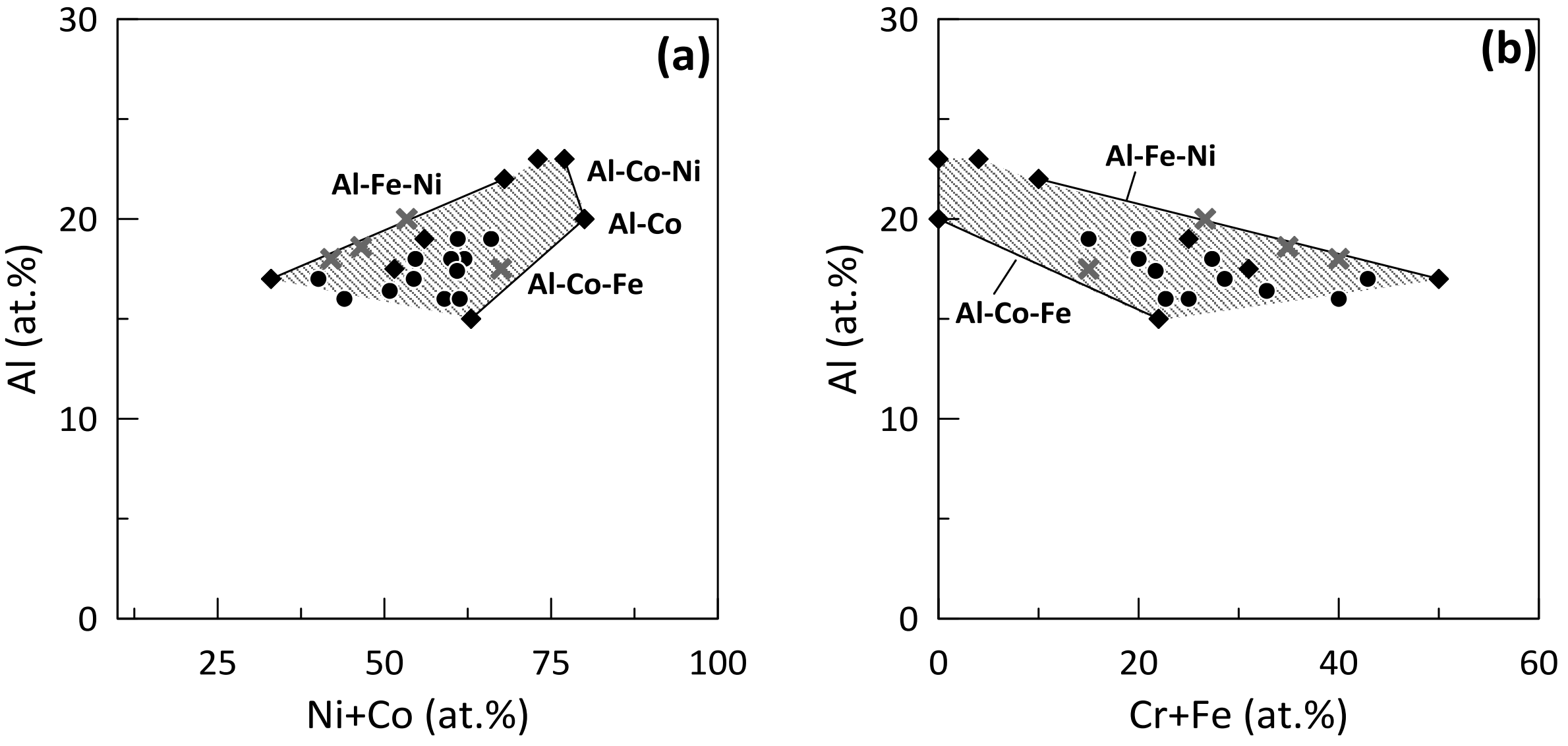

Figure 8. The Al concentrations versus (a) (Ni+Co) and (b) (Cr+Fe) concentrations of binary and ternary (♦) and quaternary and quinary (●) eutectic alloys in Table 1; four not-eutectic alloys are shown by (×)

It should be noted that every alloy which can be located inside of the eutectic regions in Figure 8 is not necessarily eutectic. To clarify this point, not-eutectic alloys in Table 2 can be considered. Most of the alloys in Table 2 will be located outside of the eutectic regions in Figure 8 (the results are not shown in Figure 8 for clarity of the diagrams), but four near-eutectic alloys $Al_{17}Co_{15}Cr_{15}Ni_{52.5}$ [12], $Al_{0.75}CoFeNi$ [35], $Al_{0.9}CrFeNi_{2.1}$ [35] and $Al_{0.8}CoCr_{0.5}FeNi$ [55] will be located within the eutectic regions close to the Al-Fe-Ni and Al-Co-Ni eutectic lines. These four not-eutectic alloys are shown in Figure 8 by crosses. Therefore, it can be concluded that every alloy which is located inside of the eutectic regions in Figure 8 is not necessarily eutectic. Eutectic regions in Figure 8 just defines the limits for the Al, (Ni+Co) and (Cr+Fe) concentrations of EHEAs.

Table 2. Chemical compositions of some of the experimentally verified not-eutectic alloys in Al-Co-Cr-Fe-Ni system

| Alloy | Chemical composition (at.%) | | | | | Eutectic phases in as cast structure |
|---|---|---|---|---|---|---|
| | Al | Co | Cr | Fe | Ni | |
| $Al_xCoCrFeNi$ [44-45] | 0 | 25 | 25 | 25 | 25 | γ (FCC) |
| $Al_xCoCrFeNi$ [44-45] | 6 | 23.5 | 23.5 | 23.5 | 23.5 | γ + B2 (BCC) |
| $Al_xCoCrFeNi$ [44-45] | 11 | 22.25 | 22.25 | 22.25 | 22.25 | γ + B2 |
| $Al_xCoCrFeNi$ [44-45] | 15 | 21.25 | 21.25 | 21.25 | 21.25 | γ + B2 |
| $Al_xCoCrFeNi$ [44-45] | 20 | 20 | 20 | 20 | 20 | α (BCC) + B2 |
| $Fe_{34}Cr_{34}Ni_{14}Al_{14}Co_4$ [46] | 14 | 4 | 34 | 34 | 14 | α + B2 |
| $Al_{0.7}NiCoFeCr_2$ [47] | 12.5 | 17.5 | 35 | 17.5 | 17.5 | α + B2 |
| $Al_{1.3}CrFeNi$ [48] | 30 | 0 | 23.33 | 23.33 | 23.34 | α + B2 |
| $Al_{12.5}Ni_{25}Co_{25}Fe_{18.75}Cr_{18.75}$ [49] | 12.5 | 25 | 18.75 | 18.75 | 25 | γ + B2 |
| $Al_{12.5}Ni_{17.5}Co_{17.5}Fe_{35}Cr_{17.5}$ [49] | 12.5 | 17.5 | 17.5 | 35 | 17.5 | α+ γ+ B2 |
| $Al_2(NiCoFeCr)_{14}$ [50] | 12.5 | 21.875 | 21.875 | 21.875 | 21.875 | α+ γ+ B2 |
| $Al_2(Ni_4Co_4Fe_3Cr_3)_{14}$ [50] | 12.5 | 25 | 18.75 | 18.75 | 25 | γ + α+B2 |
| $Al_2(NiCoFe_2Cr)_{14}$ [50] | 12.5 | 17.5 | 17.5 | 35 | 17.5 | γ + α+B2 |
| $Al_2(NiCoFeCr_2)_{14}$ [50] | 12.5 | 17.5 | 35 | 17.5 | 17.5 | γ + α+B2 |
| $Al_{0.7}NiCoFe_{1.5}Cr_{1.5}$ [51] | 12.3 | 17.5 | 26.35 | 26.35 | 17.5 | α + B2 |
| $Al_xCo_{15}Cr_{15}Ni_{70-x}$ [12] | 5 | 15 | 15 | 0 | 65 | γ |
| $Al_xCo_{15}Cr_{15}Ni_{70-x}$ [12] | 12.5 | 15 | 15 | 0 | 57.5 | γ+L12 |
| $Al_xCo_{15}Cr_{15}Ni_{70-x}$ [12] | 17.5 | 15 | 15 | 0 | 52.5 | γ+L12+B2 |
| $Al_xCo_{15}Cr_{15}Ni_{70-x}$ [12] | 27.5 | 15 | 15 | 0 | 42.5 | γ + B2 |
| $Al_xCo_{15}Cr_{15}Ni_{70-x}$ [12] | 35 | 15 | 15 | 0 | 35 | B2 |
| $Al_{0.6}CoCrFeNi$ [52] | 13 | 21.75 | 21.75 | 21.75 | 21.75 | γ + B2 |
| $Al_xCo_{2-x}CrFeNi$ [53] | 5 | 35 | 20 | 20 | 20 | γ |
| $Al_xCo_{2-x}CrFeNi$ [53] | 10 | 30 | 20 | 20 | 20 | γ + B2 |
| $Al_xCo_{2-x}CrFeNi$ [53] | 15 | 25 | 20 | 20 | 20 | γ + B2 |
| $Al_xCo_{2-x}CrFeNi$ [53] | 20 | 20 | 20 | 20 | 20 | B2 |
| $Al_xCo_{2-x}CrFeNi$ [53] | 25 | 15 | 20 | 20 | 20 | B2 |
| $Al_xCo_{2-x}CrFeNi$ [53] | 30 | 10 | 20 | 20 | 20 | B2 |
| $Al_xCo_{2-x}CrFeNi$ [53] | 35 | 5 | 20 | 20 | 20 | B2 |
| $Al_xCoFeNi$ [54] | 0 | 33.33 | 0 | 33.33 | 33.34 | γ |
| $Al_xCoFeNi$ [54] | 7.7 | 30.7 | 0 | 30.7 | 30.9 | γ |
| $Al_xCoFeNi$ [54] | 14.2 | 28.6 | 0 | 28.6 | 28.6 | γ + B2 |
| $Al_xCoFeNi$ [54] | 20 | 26.65 | 0 | 26.65 | 26.7 | γ + B2 |
| $Al_xCoFeNi$ [54] | 25 | 25 | 0 | 25 | 25 | B2 |
| $Al_xCrFeNi_{3-x}$ [35] | 20 | 0 | 20 | 20 | 40 | γ + B2 |
| $Al_xCrFeNi_{3-x}$ [35] | 18 | 0 | 20 | 20 | 42 | γ + B2 |
| $Al_xCrFeNi_{3-x}$ [35] | 14 | 0 | 20 | 20 | 46 | γ + B2 |
| $Al_xCrFeNi_{3-x}$ [35] | 12 | 0 | 20 | 20 | 48 | γ + B3 |
| $Al_xCoCr_{0.5}FeNi$ [55] | 14.6 | 24.4 | 12.2 | 24.4 | 24.4 | γ + B2 |
| $Al_xCoCr_{0.5}FeNi$ [55] | 18.6 | 23.26 | 11.63 | 23.25 | 23.26 | γ + B2 |
| $Al_xCoCr_{0.5}FeNi$ [55] | 22.23 | 22.22 | 11.11 | 22.22 | 22.22 | γ + B2 |
| $Al_xCoCr_{0.5}FeNi$ [55] | 25.53 | 21.28 | 10.63 | 21.28 | 21.28 | γ + B2 |

Figure 9a shows the Cr concentration vs (Co+Ni+Fe) concentration and Figure 9b shows the Fe concentration vs (Co+Ni+Cr) concentration of eutectic alloys in Table 1. It can be seen that the maximum concentration of Cr in eutectic alloys is 31 at.% which belongs to ternary eutectic alloy $Al_{17.5}Cr_{31}Ni_{51.5}$ [27-28]. Therefore, it can be predicted that the Cr concentration of EHEAs should be less than 31 at.%. According to Table 1 it can be seen that the Cr concentration of EHEAs is indeed less than 31 at.%. According to the diagrams in Figure 9, it can be seen that there are narrow regions limited by binary and ternary eutectic compositions in which all EHEAs are located suggesting that EHEAs are all originated from binary and ternary eutectic alloys. Furthermore, it can be anticipated that the Cr and Fe concentration of all (γ +B2) EHEAs in Al-Co-Cr-Fe-Ni system should be within the eutectic regions which are shown in Figure 9. Therefore, eutectic regions in Figure 9 can be used as references for checking the Cr and Fe concentrations of (γ +B2) EHEAs in Al-Co-Cr-Fe-Ni system. As it is shown in Figure 9, four near-eutectic alloys $Al_{17}Co_{15}Cr_{15}Ni_{52.5}$ [12], $Al_{0.75}CoFeNi$ [35], $Al_{0.9}CrFeNi_{2.1}$ [35] and $Al_{0.8}CoCr_{0.5}FeNi$ [55] are located within the eutectic regions in Figure 9. This indicates that every alloy which can be located inside of the eutectic regions in Figure 9 is not necessarily eutectic.

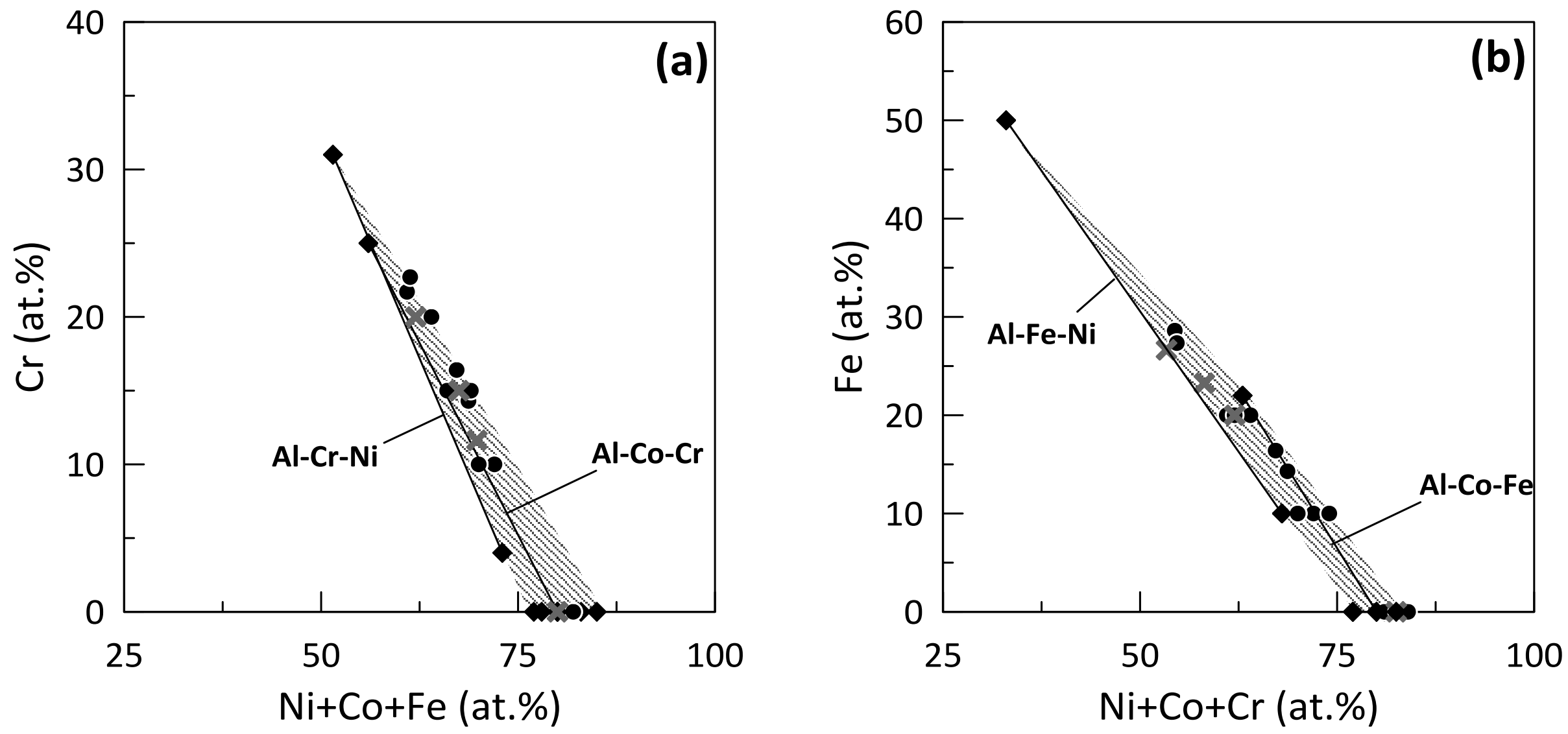

Figure 9. (a) The Cr concentrations versus (Ni+Co+Fe) concentrations and (b) the Fe concentrations versus (Ni+Co+Cr) concentrations of binary and ternary (♦) and quaternary and quinary (●) eutectic alloys in Table 1; four not-eutectic alloys are shown by (×)

Similar to diagrams in Figures 8 and 9, compositional diagrams of (Co+Ni) vs (Cr+Fe) and (Co+Cr) vs (Fe+Ni) can be considered for evaluating the composition of eutectic alloys in Al-Co-Cr-Fe-Ni system as they are shown in Figures 10a and 10b. According to these compositional diagrams, again narrow regions between binary and ternary eutectic alloys can be defined within which all EHEAs are located. Therefore, it may be speculated that all EHEAs should be within the narrow regions in Figures 10a and 10b. By considering the relation (Al+Co+Cr+Fe+Ni = 100), diagrams in Figures 10a and 10b can also be used for evaluating the Al concentration of EHEAs.

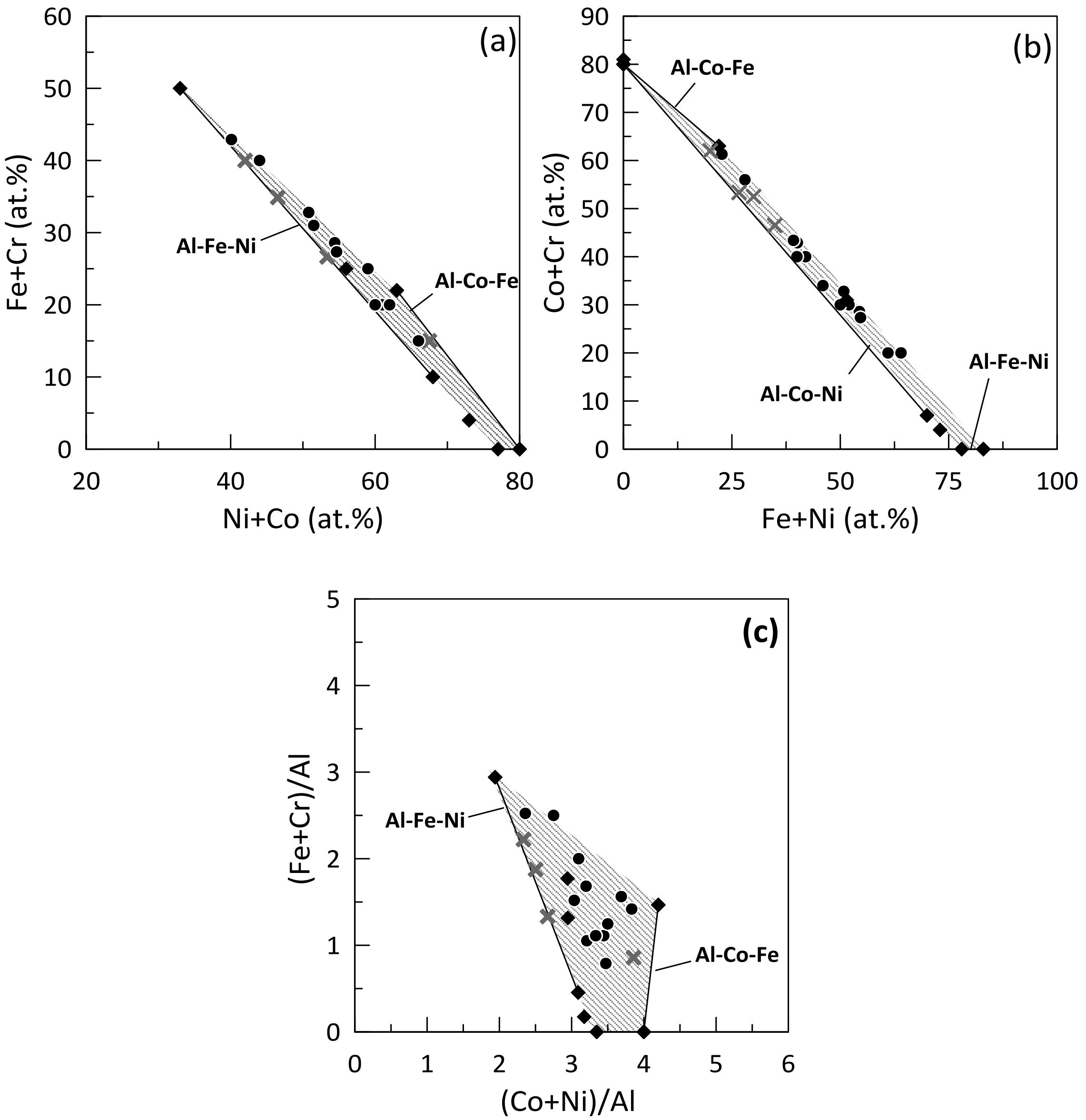

Figure 10. (a)(Co+Ni) vs (Cr+Fe), (b) (Co+Cr) vs (Ni+Fe) and (c) (Fe+Cr)/Al versus (Co+Ni)/ Al for binary and ternary (♦), and quaternary and quinary (●) eutectic alloys in Table 1; four not-eutectic alloys are shown by (×)

Another diagram for examining the compositions of EHEAs can be Al/(Co+Ni) vs Al/(Cr+Fe) which simultaneously considers the five constituent elements. This diagram is shown in Figure 10c. According to this figure, a region which is limited by binary and ternary eutectic compositions can be defined in which all EHEAs are located. The diagrams in Figure 10 may be used in alongside with diagrams in Figures 8-9 for evaluating the composition of EHEAs. According to the diagrams in Figure 10, it can be seen that four near-eutectic alloys $Al_{17}Co_{15}Cr_{15}Ni_{52.5}$ [12], $Al_{0.75}CoFeNi$ [35], $Al_{0.9}CrFeNi_{2.1}$ [35] and $Al_{0.8}CoCr_{0.5}FeNi$ [55] are located within the eutectic regions or near the eutectic region boundaries. Therefore, it can be concluded that every alloy which is located inside of the eutectic regions in Figure 10 is not necessarily eutectic. Diagrams in Figure 10 just define concentration limits for the constituent elements of EHEAs.

Diagrams in Figures 8 to 10 are in fact two dimensional projections of some parts of the quinary phase diagram Al-Co-Cr-Fe-Ni. Therefore, they cannot exactly define the eutectic volume in the quinary phase diagram Al-Co-Cr-Fe-Ni. That is why some not-eutectic alloys are also located inside of the eutectic regions in Figures 8 to 10. The diagrams in Figures 8 to 10 can only be used for evaluating the composition of EHEAs because they define the limits for the concentration of constituent elements in EHEAs.

The composition of new EHEAs cannot be extracted from eutectic regions in Figures 8 to 10. That is because not-eutectic alloys may also be located inside eutectic regions in Figures 8 to 10.  If someone wants to use these diagrams for extracting the composition of new EHEAs, then the boundaries of eutectic regions in Figures 8 to 10 should be defined more accurately which needs more experimental data. Furthermore, all of the diagrams in Figures 8 to 10 must be used simultaneously for designing new eutectic alloys. In other words, to design a new EHEA, the composition of the alloy should be checked against all of the diagrams in Figures 8 to 10 and all of the diagrams must indicate that the alloy is in eutectic regions. Furthermore, other compositional diagrams maybe developed by which eutectic and not-eutectic alloys can be categorized more accurately. If such diagrams can be developed, then they can be used for predicting the composition of EHEAs.

According to the diagrams in Figures 8 to 10, it can be seen that ranges of concentration exist for each constituent element of EHEAs in Al-Co-Cr-Fe-Ni system. Therefore, it can be concluded that a great number of $\gamma$ +B2 eutectic alloys exist in Al-Co-Cr-Fe-Ni system. Up to now, only a limited number of eutectic alloys are reported, but further eutectic alloys with excellent properties may be expected. The approach which is presented in the present work proposes a simple method for designing new eutectic alloys, and the obtained compositional diagrams present references for evaluating the composition of eutectic alloys.

## Conclusions

1- A new approach is presented for designing new eutectic high entropy alloys (EHEAs) in Al-Co-Cr-Fe-Ni system by introducing the concept of eutectic lines. It is proposed that eutectic lines exist between certain eutectic alloys in Al-Co-Cr-Fe-Ni system. As a result, new eutectic alloys can be designed by mixing the eutectic alloys which are located on the same eutectic line. By applying the proposed approach, new eutectic or near eutectic alloys are designed for Al-Co-Cr-Fe-Ni system.

2- Based on the concept of eutectic lines, a network or graph structure is proposed for eutectic alloys in Al-Co-Cr-Fe-Ni system in which the eutectic alloys which are connected via eutectic lines are determined. The proposed network can be used as a guideline for designing new eutectic alloy in Al-Co-Cr-Fe-Ni system.

3- By investigating the compositions of verified eutectic alloys in Al-Co-Cr-Fe-Ni system, compositional diagrams are proposed which show the relations between the concentrations of constituent elements in EHEAs. The proposed diagrams can be considered as convenient methods for evaluating the composition of EHEAs.

## Acknowledgments

This research was supported by Niroo Research Institute (NRI) [grant numbers 22122 & 380115]. I thank Dr. Reza Gholamipour with the Department of Advanced Materials and Renewable Energy, Iranian Research Organization for Science and Technology (IROST) for sample preparation.

## Figure captions

Figure 1. The eutectic lines for (a) Al-Co-Ni [29-30], (b) Al-Fe-Ni [26], (c) Al-Cr-Ni [27-28], (d) Al-Co-Fe, and (e) Al-Co-Cr and systems. For Al-Co-Fe system the eutectic line is drawn between the minimum eutectic composition $Al_{15}Co_{63}Fe_{22}$ [31-32] and eutectic composition $Al_{20}Co_{80}$ [34]. For Al-Co-Cr system the eutectic line is drawn according the boundaries of γ + B2 region at 1300 °C [33]

Figure 2. Proposing a network or graph structure for some of the binary, ternary, quaternary and quinary eutectic alloys in Al-Co-Cr-Fe-Ni system; each line in the network represents a eutectic line and dashed lines show eutectic lines which are not verified by experiments; as an example, the above network shows that eutectic $Al_{23}Co_7Ni_{70}$ is made by adding Ni to $Al_{20}Co_{80}$

Figure 3. The concept of eutectic lines between eutectic alloys in quinary Al-Co-Cr-Fe-Ni system

Figure 4. (a) Assuming eutectic lines between eutectic alloys $Al_{20}Co_{80}$ and $Al_{19}Cr_{15}Co_{15}Ni_{51}$ [12] and between $Al_{23}Co_7Ni_{70}$ [30] and $Al_{19}Co_{15}Cr_{15}Ni_{51}$ [12]; obtaining new eutectic alloys with compositions $Al_{19.5}Co_{47.5}Cr_{7.5}Ni_{25.5}$ and $Al_{21}Co_{11}Cr_{7.5}Ni_{60.5}$; the simulation results for alloys (b) $Al_{19.5}Co_{47.5}Cr_{7.5}Ni_{25.5}$ and (c) $Al_{21}Co_{11}Cr_{7.5}Ni_{60.5}$

Figure 5. Optical images of the as-cast microstructure of alloy $Al_{19.5}Co_{47.5}Cr_{7.5}Ni_{25.5}$

Figure 6. The eutectic network obtained by considering some of the combinations of eutectic alloys in Table 1; the narrow eutectic lines are verified by simulation results, and the bold eutectic lines are verified by experiments, the dotted lines show the lines which are investigated but are not eutectic according to simulation results; the simulation results show that alloys along the dotted lines are in fact near eutectic alloys

Figure 7. Simulation results for alloy $Al_{19}Cr_{10}Fe_{15}Ni_{56}$ obtained by mixing eutectic alloys $Al_{22}Fe_{10}Ni_{68}$ [26] and $AlCrFeNi_2$ [35], (b) assuming a curved eutectic line between alloys $Al_{22}Fe_{10}Ni_{68}$ [26] and $AlCrFeNi_2$ [35]; in this condition, alloys cannot be mixed with each other for designing new eutectic alloys

Figure 8. The Al concentrations versus (a) (Ni+Co) and (b) (Cr+Fe) concentrations of binary and ternary (♦) and quaternary and quinary (●) eutectic alloys in Table 1; four not-eutectic alloys are shown by (×)

Figure 9. (a) The Cr concentrations versus (Ni+Co+Fe) concentrations and (b) the Fe concentrations versus (Ni+Co+Cr) concentrations of binary and ternary (♦) and quaternary and quinary (●) eutectic alloys in Table 1; four not-eutectic alloys are shown by (×)

Figure 10. (a)(Co+Ni) vs (Cr+Fe), (b) (Co+Cr) vs (Ni+Fe) and (c) (Fe+Cr)/Al versus (Co+Ni)/ Al for binary and ternary (♦), and quaternary and quinary (●) eutectic alloys in Table 1; four not-eutectic alloys are shown by (×)